\documentclass[11pt,a4paper]{article}

\if0
\documentclass[12pt, a4paper]{article}
\usepackage{setspace}
\fi

\usepackage[round]{natbib}

\usepackage{amsmath}
\usepackage{ascmac}
\usepackage{amsfonts}
\usepackage[dvipdfmx]{graphicx}
\usepackage{url} %%url

\makeatletter
  \let\c@figure\c@table
  \let\p@figure\p@table
  \let\cl@figure\cl@table
  
\makeatother

\makeatletter
   
   \@addtoreset{table}{section}
\makeatother

   

\newcommand{\dd}{{\mathrm d}}

\newcommand{\ddef}{\stackrel{{\rm def}}{=}}
\newcommand{\argmax}{\mathop{\rm argmax}\limits}

\newcommand{\E}{\mathrm{E}}
\newtheorem{theorem}{Theorem}[section]
\newtheorem{definition}[theorem]{Definition}
\newtheorem{proposition}[theorem]{Proposition}

\newtheorem{example}[theorem]{Example}

\title{Determinantal Point Process Priors \\for Bayesian Variable Selection \\ in Linear Regression}
\author{Mutsuki KOJIMA${}^*$ and Fumiyasu KOMAKI${}^*{}^{\dagger}$\\ \\
        ${}^*$Department of Mathematical Informatics\\
        Graduate School of Information Science and Technology\\
        The University of Tokyo, Tokyo, Japan\\
        ${}^{\dagger}$RIKEN Brain Science Institute, Wako-shi, Japan\\
        \texttt{\normalsize \{mutsuki\_kojima,komaki\}@mist.i.u-tokyo.ac.jp}}
\date{}

\begin{document}
\maketitle

\begin{abstract}
We propose discrete determinantal point processes (DPPs) for priors on the model parameter in Bayesian variable selection. By our variable selection method, collinear predictors are less likely to be selected simultaneously because of the repulsion property of discrete DPPs. Three types of DPP priors are proposed. We show the efficiency of the proposed priors through numerical experiments and applications to collinear datasets.
\end{abstract}

\section{Introduction}
We consider Bayesian variable selection in linear regression. Suppose we have $n$ observations on a dependent variable $y_n~(n\times 1~{\rm matrix})$ and $p$ predictor variables $X=(x_1,\ldots,x_p)~(n\times p~{\rm matrix})$, for which the normal linear model holds:
\begin{eqnarray}\label{full_model}
y_n=X\beta+\varepsilon_n,
\end{eqnarray}
where $\varepsilon_n \sim \mathcal{N}(0, \sigma^2I_n)~(n\times 1~{\rm matrix})$ and  $\beta=(\beta_1,\ldots, \beta_p)^{\top}~(p\times 1~{\rm matrix})$. Let $\gamma = (\gamma_1,\ldots, \gamma_p)^{\top} \in \{0,1\}^p$ be a model parameter, meaning that $\gamma_i=1$ indicates $\beta_i$ is nonzero and $\gamma_i=0$ indicates $\beta_i=0$. For Bayesian variable selection, we consider $2^p$ possible submodels of \eqref{full_model}. Each submodel is denoted by $M_{\gamma}$. Let $X_{\gamma}$ be a $n\times |\gamma|$ design matrix consisting of these columns of $X$ that correspond to the predictors with $\gamma_i=1$. Here, $|\gamma|$ is the number of nonzero elements of $\gamma$, i.e., $|\gamma|=\sum_{i=1}^p\gamma_i$. Under submodel $M_\gamma$, $y_n$ follows
\begin{eqnarray}
M_{\gamma}:\quad y_n=X_{\gamma}\beta_{\gamma} + \varepsilon_n,
\end{eqnarray}
where $\beta_{\gamma}$ is the $|\gamma|$-dimensional vector of nonzero regression coefficients of $\beta$ with $\gamma_i=1$. Bayesian variable selection is to identify nonzero components of $\beta$ assigning
priors to the parameters. We select the best model that attains the maximum of the posterior probability $p(\gamma|y_n)$.

The normal linear regression model is simple and useful, but collinearity problem often arises when we apply it to a real data. A serious problem of collinearity is imprecision of the ordinary least squares estimator (OLS). The problem occurs when $X$ is ill-conditioned, i.e., highly correlated predictors exist. Then $(X^{\top}X)^{-1}$ is numerically unstable and therefore the OLS, $\hat{\beta}_{\rm OLS}=(X^{\top}X)^{-1}X^{\top}y_n$, is not reliable. One way of avoiding the issue is variable selection. However, most existing methods for variable selection do not take into account for the correlations among predictors. In the large $p$ small $n$ setting, in which there are many more predictors than observations, \cite{KwonEtAl} exploits the correlations for proposal distribution in a stochastic search method. However,  few Bayesian variable selection methods considering the correlations have been proposed. We propose discrete determinantal point processes (DPPs) for the prior distribution on $\gamma$, so that submodels including collinear predictors are less likely to be selected. Our Bayesian variable selection method makes use of  the correlations among predictors. 

DPPs have been studied since \cite{Macchi} first identified them as a class of point processes. Recently, \cite{BorodinRains} introduced discrete DPPs, and discrete DPPs have been applied to machine learning problems by \cite{KuleszaTaskar}. Discrete DPPs are elegant probabilistic models of repulsion, and \cite{KuleszaTaskar} considered repulsion as diversity of items. For example, in a document summarization task,  modeling the task with discrete DPPs is appropriate because a summary requires diversity and quality.

Selected predictors in variable selection should be diverse, which means each pair of selected predictors is nearly uncorrelated. For the selection of diverse predictors, discrete DPPs are efficient priors on $\gamma$ in Bayesian variable selection. We show that discrete DPPs are useful priors through numerical experiments and applications to real data sets.

The remainder of this paper is organized as follows. In Section 2, the definition and examples of discrete DPPs are given. In Section 3, we first review the Bayesian variable selection method proposed by \cite{GeorgeFoster}. Next, we propose Bayesian variable selection methods using three types of DPP priors on $\gamma$. In Section 4, we show numerical experimental results and we report applications to the two real datasets in Section 5. We conclude the paper in Section 6.

\section{Discrete Determinantal Point Processes}\label{dpp}
First, we give the definition of discrete determinatal point processes (DPPs). 

Let $\Lambda$ be $\{1,\ldots, p\}$ and let $L$ be a $p\times p$ symmetric positive definite matrix. We identify $\{0,1\}^p$ with the power set of $\Lambda$ ($2^{\Lambda}$). To be precise, for $\gamma\in\{0,1\}^p$, $\gamma_i=1$ indicates $i\in \gamma$ and $\gamma_i=0$ indicates $i\not\in\gamma$.

\begin{definition}\label{def_dpp}
\upshape
 A random variable $\mathcal{X}$ that takes values in the power set of $\Lambda$ is called {\it discrete determinantal point process (DPP) with kernel} $L$, if 
\begin{eqnarray}
P(\mathcal{X}=\gamma)\propto \det (L_{\gamma}),
\end{eqnarray}
where $\gamma \in \{0,1\}^p$ and $L_{\gamma}$ is the $|\gamma|\times |\gamma|$ matrix whose elements are $(L_{ij})_{i,j\in\gamma}$. We define $\det(L_{\emptyset})=1$.
\end{definition}

The normalization constant is given by the next proposition. See \cite{KuleszaTaskar} for the proof.

\begin{proposition}\label{norm_const}
\upshape
\begin{eqnarray}
\sum_{\gamma\in \{0,1\}^p} \det(L_{\gamma})=\det(L+I_p),
\end{eqnarray}
where $I_p$ is the $p\times p$ identity matrix, and the sum is taken over all subsets of $\Lambda$.
\end{proposition} 

Definition \ref{def_dpp} and Proposition \ref{norm_const} are enough to propose DPP priors. For more detail properties of discrete DPPs, see \cite{KuleszaTaskar}. See \cite{HoughEtAl}, for general determinantal point processes.

Next, we give a brief explanation of the repulsion property of DPPs. The property is a key of our proposal. Let $F$ be a $p\times q~(p<q)$ matrix, and we denote the rows of $F$ by $f_i~(i=1,2,\ldots, p)$. We assume that $L=FF^{\top}$. If $\mathcal{X}$ follows the discrete DPP with kernel $L$, then
\begin{eqnarray}\label{geometry_of_dpp}
P(\mathcal{X}=\gamma)\propto ({\rm vol}(\{f_i\}_{i\in\gamma}))^2,
\end{eqnarray}
where ${\rm vol}(\{f_i\}_{i\in\gamma})$ means the $|\gamma|$-dimensional volume of the parallelepiped spanned by the rows of $\{f_i\}_{i\in\gamma}$. By considering $f_i$ as the feature vector of item $i$, ${\rm vol}(\{f_i\}_{i\in\gamma})$ is small if there exist similar items in $\gamma$. This is because the volume of the parallelepiped is smaller as the angle between two edges decreases. Since the small angle between edges $f_i$ and $f_j$ means that item $i$ is similar to item $j$, the probabilities that $\mathcal{X}$ includes similar items are small. Thus DPPs are considered to prefer repulsion and diversity. 

Finally we give two examples of discrete DPPs. The first example shows the distribution of  discrete DPPs with a diagonal matrix kernel corresponds to the Bernoulli distribution. The second example indicates that nonzero off-diagonal elements of the kernel $L$ determine negative correlations between pairs of items. 

\begin{example}\label{dpp_diag_kernel}
\upshape
Let $L=(L_{ij})_{i,j=1,\ldots,p}$ be a diagonal matrix whose elements are
\begin{eqnarray}
L_{ij}=\left\{
\begin{array}{ll}
w/(1-w) & {\rm if}~ i=j,\\
0 & {\rm otherwise},
\end{array}
\right.
\end{eqnarray}
where $w \in (0,1)$. Suppose $\mathcal{X}$ follows the discrete DPPs with kernel $L$, then by proposition \ref{norm_const}, 
\[
P(X=\gamma)=\frac{\det(L_{\gamma})}{\det(L+I_p)}=w^{|\gamma|}(1-w)^{p-|\gamma|}.
\]
In this setting, the distribution of $\mathcal{X}$ is just the Bernoulli distribution with success probability $w$.
\end{example}

\begin{example}\label{off_diag_dpp}
\upshape
Let $\Lambda=\{1,2,3\}$ and 
\begin{eqnarray*}
L=
\begin{pmatrix}
1 & 0.9 & 0\\
0.9 & 1 & 0\\
0 & 0 & 1
\end{pmatrix}
.
\end{eqnarray*}
Suppose $\mathcal{X}$ follows the discrete DPPs with kernel $L$. The distribution of $\mathcal{X}$ is as Table \ref{ex2_dpp}. From Table \ref{ex2_dpp}, the event that $\mathcal{X}$ equals $\{1,2\}$ or $\{1, 2, 3\}$ is less likely to occur. This is because the off-diagonal element $L_{12}=0.9$. In fact, 
\begin{eqnarray}
P(\mathcal{X}=\{i, j\})\propto P(\mathcal{X}=\{i\})P(\mathcal{X}=\{j\})-\bigg(\frac{L_{ij}}{\det(L+I_p)}\bigg)^2.
\end{eqnarray}
\end{example}
Thus, if off-diagonal elements are nonzero, then the corresponding items are less likely to be included simultaneously.

\begin{table}[tb]
\begin{center}
\caption{The distribution of  $\mathcal{X}$ in Example \ref{off_diag_dpp}.}

\begin{tabular}{|c|c|}\hline
\label{ex2_dpp}
Subsets & Probabilities\\ \hline
$\emptyset$ & 0.157\\
$\{1\}$ & 0.157\\
$\{2\}$ & 0.157\\
$\{3\}$ & 0.157\\\hline
\end{tabular}
\begin{tabular}{|c|c|} \hline
Subsets & Probabilities\\ \hline
$\{1, 2\}$ & 0.030\\
$\{1, 3\}$ & 0.157\\
$\{2, 3\}$ & 0.157\\
$\{1, 2, 3\}$& 0.030\\ \hline
\end{tabular}
\end{center}
\end{table}

\section{Bayesian Variable Selection Methods}\label{bayesian_variable_selection}
In subsection \ref{GeorgeFosterMethod}, we first review the Bayesian variable selection method proposed by \cite{GeorgeFoster}. In subsection \ref{proposedMethods}, we propose Bayesian variable selection methods using three types of DPP priors on $\gamma$.

\subsection{Bayesian Variable Selection Method Proposed by \cite{GeorgeFoster}}\label{GeorgeFosterMethod}

\cite{GeorgeFoster} proposed the following Bayesian variable selection.

Zellner's $g$-prior \citep{Zellner} is assigned to nonzero regression coefficients $\beta_\gamma$ under submodel $M_{\gamma}$:
\begin{eqnarray}\label{g_prior}
p(\beta_{\gamma}|g)\sim \mathcal{N}(0, g\sigma^2(X_{\gamma}^{\top}X_{\gamma})^{-1}),\quad g>0,
\end{eqnarray}
where $g$ is the hyperparameter. For the prior distribution on model parameter $\gamma$, the Bernoulli distribution
\begin{eqnarray}
p(\gamma|w)=w^{|\gamma|}(1-w)^{p-|\gamma|}
\end{eqnarray}
with success parameter $w\in (0,1)$ is used. The best model
\begin{eqnarray}
\hat{\gamma}&=&\argmax_{\gamma}~p(\gamma|y_n, g,w)\nonumber\\
&=&\argmax_{\gamma}~\exp \bigg(\frac{g}{2(1+g)}({\rm ss}_{\gamma}/\sigma^2-F(g,w)|\gamma|)\bigg)\nonumber\\
&=&\argmax_{\gamma}~ ({\rm ss}_{\gamma}/\sigma^2-F(g,w)|\gamma|), \label{InfoCriteria}
\end{eqnarray}
maximizing the posterior probability, is selected, where
\begin{eqnarray}
{\rm ss}_{\gamma}=y_n^{\top}X_{\gamma}X_{\gamma}^{\top}y_n,\quad F(g, w)=\frac{1+g}{g}\bigg(2\log\frac{1-w}{w}+\log(1+g)\bigg).
\end{eqnarray}
If we assume $\sigma^2$ is known and hyperparameters are appropriately calibrated, the Bayesian variable selection above is identical to selecting the best model by the typical penalized sum of squares criteria, such as AIC \citep{Akaike}, BIC \citep{Schwarz} or RIC \citep{FosterGeorge}. For example, if we set $g$ and $w$ such that $F(g, w)=2$, then the highest posterior model is the model maximizing \eqref{InfoCriteria}
\begin{eqnarray}
{\rm ss}_{\gamma}/\sigma^2-2|\gamma|.
\end{eqnarray}
In this setting, the highest posterior model exactly corresponds to the best model with the lowest AIC. 

For hyperparameters $g$ and $w$, \cite{GeorgeFoster} used type II maximum likelihood estimators $\hat{g}$ and $\hat{w}$ given by
\begin{eqnarray}
(\hat{g}, \hat{w})&=& \argmax_{g,w}~p(y_n|g,w)\nonumber\\
&=&\argmax_{g,w} \sum_{\gamma\in\{0,1\}^p}p(\gamma|w)\int p(y_n|\gamma, \beta_{\gamma})p(\beta_{\gamma}|g)\dd \beta_{\gamma}.
\end{eqnarray}
Since Zellner's $g$-prior is the normal distribution, the marginal distribution can is represented in the closed-form:
\begin{eqnarray}
\int p(y_n|\gamma, \beta_{\gamma})p(\beta_{\gamma}|g)\dd \beta_{\gamma}= \frac{(1+g)^{-|\gamma|/2}}{(2\pi)^{n/2}\sigma^n}\exp\bigg(\frac{g}{1+g}\frac{{\rm ss}_{\gamma}}{2\sigma^2}-\frac{y_n^{\top}y_n}{2\sigma^2}\bigg).
\end{eqnarray}
Therefore the type II likelihood for $g$ and $w$ is 
\begin{eqnarray}
p(y_n|g,w)\propto \sum_{\gamma\in\{0,1\}^p} \frac{w^{|\gamma|}(1-w)^{p-|\gamma|}}{\sigma^n(1+g)^{|\gamma|/2}}\exp\bigg(\frac{g}{1+g}\frac{{\rm ss}_{\gamma}}{2\sigma^2}-\frac{y_n^{\top}y_n}{2\sigma^2}\bigg).
\end{eqnarray}
We refer the method above as EB (empirical Bayes) in the following sections.

\subsection{DPP Priors and Proposed Methods}\label{proposedMethods}
Let $x_{ij}$ be the $(i,j)$ element of design matrix $X$ and let $\tilde{X}=(\tilde{x}_1, \ldots, \tilde{x}_p)$ ($n\times p$ matrix) be the standardized matrix of design matrix $X$. To be precise, the $(i,j)$ element $\tilde{x}_{ij}$ of $\tilde{X}$ is defined by
\begin{eqnarray*}
\tilde{x}_{ij}=\frac{x_{ij}-m_j}{s_j},
\end{eqnarray*}
where
\begin{eqnarray*}
m_j\ddef \frac{1}{n}\sum_{i=1}^nx_{ij}, \quad s_j\ddef \sqrt{\sum_{i=1}^n(x_{ij}-m_j)^2}.
\end{eqnarray*}
We denote the correlation matrix ($p\times p$ matrix) of $X$ by $R$ given by
\begin{eqnarray}
R=\tilde{X}^{\top}\tilde{X}.
\end{eqnarray}
The first proposal for prior distribution $p(\gamma|w)$ is
\begin{eqnarray}\label{dpp_cor_ker_prior}
p(\gamma|w)\propto \det (wR_{\gamma})=w^{|\gamma|}\det(R_{\gamma}), \quad w>0.
\end{eqnarray}
By Proposition \ref{norm_const}, 
\begin{eqnarray}
p(\gamma|w)=\frac{\det(wR_{\gamma})}{\det(wR+I_p)}.
\end{eqnarray}
We call this prior {\it DPP prior} (referred as DPP). This proposal is based on the following consideration. Assume there are three predictors $x_1$,  $x_2$ and $x_3$. We also assume that the correlation between $x_1$ and $x_2$ is $0.9$ and the other pairs are uncorrelated. If $w=1$, then the probabilities of $p(\gamma|w)$ are the same as Table \ref{ex2_dpp}.
From Table \ref{ex2_dpp}, we see that DPP prior assigns small probabilities to subsets of predictors including collinear predictors ($x_1$ and $x_2$). DPPs prefer repulsion and diversity as we see in Section \ref{dpp}. Therefore, when we put DPP prior on model parameter $\gamma$, we can select diverse predictors, which means each pair of selected predictors is nearly uncorrelated. The hyperparameter $w$ controls the expected proportion of nonzero regression coefficients; if $w>1$ then larger subsets are more preferable, otherwise smaller subsets are more preferable.

From Example \ref{dpp_diag_kernel}, the Bernoulli distribution is a discrete DPP with a diagonal matrix kernel. Thus DPP prior is a generalization of the Bernoulli distribution that is used for $p(\gamma|w)$ in EB. From this point of view, we propose two types of priors that bridge the Bernoulli distribution and DPP prior:
\begin{eqnarray}
\label{linear_mix_dpp}
p(\gamma|w,\theta)&\propto & \det(w(\theta R_{\gamma} + (1-\theta)I_{\gamma})),\quad w>0,\quad \theta\in [0,1],\\
\label{geo_mix_dpp}
p(\gamma|w,\alpha)&\propto & \det(w(R^{\alpha})_{\gamma}), \quad w>0, \quad \alpha \geq 0,
\end{eqnarray}
where $I_{\gamma}$ is the $|\gamma|\times |\gamma|$ identity matrix and $R^{\alpha}$ is the non-integer powers $\alpha$ of $R$. We call \eqref{linear_mix_dpp} \textit{linear mixture DPP prior} (referred as LDPP) and \eqref{geo_mix_dpp} \textit{geometric mixture DPP prior} (referred as GDPP). Linear mixture DPP prior corresponds to DPP prior when $\theta=1$ and corresponds to the Bernoulli distribution when $\theta=0$. Similarly, geometric mixture DPP prior corresponds to DPP prior when $\alpha=1$ and corresponds to the Bernoulli distribution when $\alpha=0$. 

Our Bayesian variable selection methods are as follows. We put proposed priors (DPP, LDPP or GDPP) on model parameter $\gamma$ and $g$-prior on $\beta_{\gamma}$. Next, hyperparameters are estimated by maximizing the type II likelihood. Here hyperparameters are $g$, $\sigma^2$ (if not known),  $w$, $\theta$ (if using LDPP) and $\alpha$ (if using GDPP).  The best model is selected that maximizes the posterior probability $p(\gamma | y_n)$.

Though we discuss variable selection methods so far, we can estimate the regression coefficients after selecting the best model. The estimator $\hat{\beta}$ is constructed after estimating $\hat{g}$ and selecting the best model $M_{\hat{\gamma}}$:
\begin{eqnarray}\label{reg_coeff_estimator}
\hat{\beta}=\E [\beta|\hat{\gamma},\hat{g}]=\frac{\hat{g}}{1+\hat{g}}(X_{\hat{\gamma}}^{\top}X_{\hat{\gamma}})^{-1}X_{\hat{\gamma}}^{\top}y_n.
\end{eqnarray}
The representation of the estimator is the same whether the best model $\hat{\gamma}$ is selected by EB or by our methods. This is because the $g$-prior is put on the regression coefficients in both methods.

\section{Numerical Experiments}
In this section, we evaluate the risk of estimated regression coefficients as the sample size increases through numerical experiments. 

In the numerical experiments, the following settings are considered. First we sample a $400\times 6$ design matrix $X^*$ whose columns are $x_i^*~(i=1,2,\ldots, 6)$:
\begin{eqnarray*}
x_1^*,~x_2^*,~x_3^*,~\varepsilon_4,~\varepsilon_5,~\varepsilon_6~&\stackrel{\rm i.i.d.}{\sim}& \mathcal{N}_{400}(0, I_{400}),\\
x_4^* &=& x_1^* + x_2^* + 0.1\times \varepsilon_4,\\
x_5^* &=& x_1^* + x_3^* + 0.1\times \varepsilon_5,\\
x_6^* &=& x_2^* + x_3^* + 0.1\times \varepsilon_6.
\end{eqnarray*} 
Let $X_k^*~(k=1,2,\ldots,20)$ be a $20k \times 6$ submatrix of $X^*$ whose rows correspond to the first $20k$ rows of $X^*$ and $\beta^*=(1,-1,0,0,0,0)^{\top}$. For each $k$, we simulate $y_{20k}$ 10000 times following \eqref{full_model} with $X=X_k^*$, $\beta=\beta^*$ and $\sigma^2=0.9^2$. For each $y_{20k}$, the estimator $\hat{\beta}$ is constructed by each method. A loss of each estimator is averaged over 10000 repetitions. 

For the loss function, we employ the maximum loss function $\|\beta^*-\hat{\beta}\|_{\infty}$ defined by
\begin{eqnarray}
\|\beta^*-\hat{\beta}\|_{\infty}\ddef \max_i |\beta^*_i-\hat{\beta}_i|,
\end{eqnarray}
where $\hat{\beta}=(\hat{\beta}_1,\ldots ,\hat{\beta}_p)^{\top}$. A usual loss function for estimators of regression coefficients is quadratic loss (i.e., $\|\beta^*-\hat{\beta}\|_2^2=\sum_i |\beta^*_i-\hat{\beta}_i|^2$) or predictive loss (i.e., $\|X\beta^*-X\hat{\beta}\|_2^2=\sum_i |x_i\beta^*_i-x_i\hat{\beta}_i|^2$). However, the maximum loss function is more appropriate than these usual loss functions when influence of collinearity on the estimated regression coefficients is investigated. The reason is described in the last paragraph of this section. Therefore we use the maximum loss function in the numerical experiments.

We denote $\hat{\beta}_{\rm EB}$, $\hat{\beta}_{\rm DPP}$, $\hat{\beta}_{\rm LDPP}$ and $\hat{\beta}_{\rm GDPP}$ defined by \eqref{reg_coeff_estimator} when for $p(\gamma|w)$ using the Bernoulli distribution (EB), DPP prior (DPP), linear mixture DPP prior (LDPP), and geometric mixture DPP prior (GDPP), respectively. We assume that $\sigma^2=0.9^2$ is known and estimate other hyperparameters $g$, $w$, $\theta$ (in LDPP)  by maximizing the type II likelihood. For example, when using LDPP, we estimate $g$, $w$ and $\theta$ by maximizing the marginal likelihood
\begin{eqnarray}
p(y_n|g,w,\theta)&\propto& \sum_{\gamma\in\{0,1\}^p}\bigg\{ \frac{\det (w(\theta R_{\gamma} + (1-\theta)I_{\gamma}))}{(1+g)^{|\gamma|/2}\det (w\theta R + (1+w-w\theta)I)}\nonumber\\
&&\quad \quad \quad \times \exp\bigg(\frac{g}{1+g}\frac{{\rm ss}_{\gamma}}{2\sigma^2}\bigg)\bigg\}.
\end{eqnarray}
For $\alpha$ (in GDPP), we use the parameter $\hat{\alpha}$ that maximizes the type II likelihood $p(y_n|\alpha)$ over $[0, 3]$. Since $X^{\top}X$ is ill-conditioned in the numerical experiments, we restrict the domain of the optimization.

For comparison, other estimators are also investigated: 
\begin{eqnarray}
\hat{\beta}_{\rm RIDGE}&\ddef&(X^{\top}X+\lambda I_p)^{-1}X^{\top}y_n,\\
\hat{\beta}_{\rm OLS}&\ddef&(X^{\top}X)^{-1}X^{\top}y_n,\\
\hat{\beta}_{\rm ORACLE}&\ddef&(X_{\gamma^*}^{\top}X_{\gamma^*})^{-1}X_{\gamma^*}^{\top}y_n,
\end{eqnarray}
where $\lambda >0$ is the hyperparameter in ridge regression (putting $\mathcal{N} (0, \sigma^2\lambda^{-1} I)$ on $\beta$) and $\gamma^{*}$ is the true subset of nonzero coefficients. We estimate $\lambda$ by maximizing the type II likelihood
\begin{eqnarray*}
p(y_n|\lambda)&=&\int p(y_n|\beta)p(\beta|\lambda)\dd\beta\\
&\propto&\lambda^{p/2}\int \exp\bigg(-\frac{1}{2\sigma^2}(y_n-X\beta)^{\top}(y_n-X\beta)-\frac{\lambda}{2\sigma^2}(\beta^{\top}\beta)\bigg)\dd\beta.
\end{eqnarray*}

\begin{figure}[htbp]
\centering
\includegraphics[width=8.5cm, height=8.5cm]{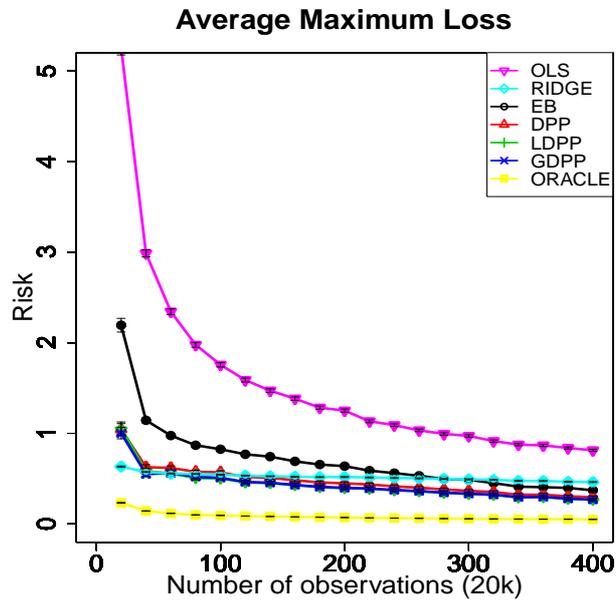}
\includegraphics[width=8.5cm, height=8.5cm]{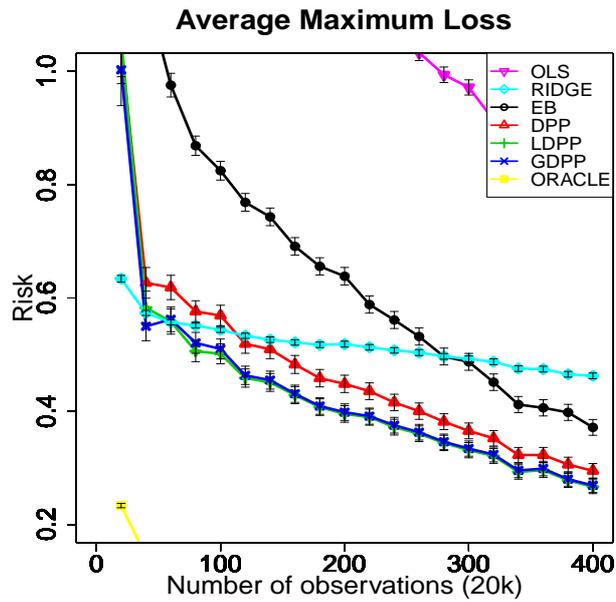}
\caption{Comparison of the maximum risk for each procedure (EB, DPP, LDPP, GDPP, RIDGE, OLS) as the sample size increases: The upper panel shows the result of estimated the maximum risk for each procedure. Each point displays the average maximum risk at $20k~(k=1,2,\ldots, 20)$ observations. Error bars indicate mean $\pm 3\times$standard error.  The bottom panel shows an enlargement of the upper panel.}
\label{result_experiments}
\end{figure}

Figure \ref{result_experiments} shows the results of the comparison. From Figure \ref{result_experiments}, DPP, LDPP and GDPP outperform the other estimators. In Section \ref{bayesian_variable_selection}, we see that the best model selected by EB corresponds to the best model selected by the typical penalized sum of squares criteria, such as AIC, BIC or RIC. Therefore EB is considered to evaluate complexity of a submodel by its dimension. Since DPP, LDPP and GDPP penalize a submodel not only by its dimension but also by the correlations among predictors $X$, they perform better than EB. Though RIDGE can reduce the quadratic loss, but its maximum risk is worse than DPP, LDPP and GDPP.

Finally we give the reason why the maximum loss is more appropriate than the usual loss functions. A serious problem of collinearity is the imprecision of OLS. It is well known that the predictive loss is not affected by the collinearity even if collinearity is severe. This is because specific combinations of estimated regression coefficients are well-determined by the ordinary least squares \citep{BelsleyKuhWelsch}. Therefore the predictive loss is not appropriate for investigation of the influence of collinearity on estimated regression coefficients. Since the quadratic loss is mathematically tractable, it has been used when dealing with correlated predictors. Ridge regression \citep{HoerlKennard} is the method of constructing estimators with the quadratic penalty for estimated coefficients. However, the quadratic loss may be inappropriate since it summarizes componentwise distances from an estimator to the true parameter. If we want to estimate the all values of regression coefficients precisely, the maximum loss is more appropriate than the quadratic loss since the maximum loss evaluates the furthest distance in all components. For example, assume that three predictors $\{x_i\}_{i=1}^3$ exist and $x_3$ nearly equals to $x_1+x_2$. Assume also that the true regression coefficients are $\beta^*=(\beta_1^*, \beta_2^*, \beta_3^*)^{\top}=(1,-1,0)^{\top}$ and two estimators $\hat{\beta}^{(1)}$ and $\hat{\beta}^{(2)}$ are obtained:
\begin{eqnarray}
\hat{\beta}^{(1)}=(0.5,-1.5,0.0)^{\top},\quad \hat{\beta}^{(2)}=(1.1,-0.9,-0.6)^{\top}.
\end{eqnarray}
$\hat{\beta}^{(1)}$ is more favorable than $\hat{\beta}^{(2)}$ since the third component of  $\hat{\beta}^{(1)}$ is zero but that of $\hat{\beta}^{(2)}$ is nonzero. However, each quadratic loss is
\begin{eqnarray}
\|\beta^*-\hat{\beta}^{(1)}\|_2^2=0.5,\quad \|\beta^*-\hat{\beta}^{(2)}\|_2^2=0.38.
\end{eqnarray}
This means $\hat{\beta}^{(2)}$ is more preferable than $\hat{\beta}^{(1)}$. In contrast, each maximum loss is
\begin{eqnarray}
\|\beta^*-\hat{\beta}^{(1)}\|_{\infty}=0.5,\quad \|\beta^*-\hat{\beta}^{(2)}\|_{\infty}=0.6,
\end{eqnarray}
which means $\hat{\beta}^{(1)}$ is more preferable.  Even though the third component of $\hat{\beta}^{(2)}$ is far from the true parameter $\beta_3^*$, the distances of the other components from $\hat{\beta}^{(2)}$ to the true parameter are closer than those of $\hat{\beta}^{(1)}$. Therefore, in total, $\hat{\beta}^{(2)}$ is more preferable than $\hat{\beta}^{(1)}$ with respect to the quadratic loss. This happens because the quadratic loss function summarizes componentwise distances from an estimator to the true parameter. For this reason, the maximum loss is more appropriate than the quadratic loss when we are interested in the impact of collinearity on all components of estimated regression coefficients.

\section{Applications to Real Datasets}
Let $x^k$ be the $k$-th row of the design matrix $X$. In this section, we call $x^k$ the $k$-th observation. Note that $x_i$ is the $i$-th predictor.

In this section, we report the results of applications to the Air Pollution Data and the Body Fat Data. Before showing the result, we summarize assumptions and analysis methods for the datasets. 

In practice, since the mean of the dependent variable $y_n$ is almost always nonzero, we assume that the constant term $1_n=(1,\ldots, 1)^{\top}$ ($n\times 1$ matrix) is included. Therefore we consider the following linear regression model:
\begin{eqnarray}
y_n=\mu 1_n + X\beta +\varepsilon_n,
\end{eqnarray}
where $\mu $ is an unknown intercept parameter. In Bayesian variable selection, we consider the following submodels:
\begin{eqnarray}
M_{\gamma}: y_n=\mu 1_n+X_{\gamma}\beta_{\gamma}+\varepsilon_n.
\end{eqnarray}
We also assume that the columns of $X$ is standardized. 

We compare proposed methods and the usual methods (EB, RIDGE and OLS). For Bayesian variable selection (EB, DPP, LDPP and GDPP), hyperparameters $\mu$, $\sigma^2$, $g$, $w$, $\theta$ (if using LDPP) are estimated by maximizing the type II likelihood. For $\alpha$ (in GDPP), we use the parameter $\hat{\alpha}$ that maximizes the type II likelihood $p(y_n|\alpha)$ over $[0, 3]$ since $X^{\top}X$ is ill-conditioned. The best model is selected that maximizes the posterior probability $p(\gamma | y_n)$. The estimator of regression coefficients $\hat{\beta}$ is constructed following \eqref{reg_coeff_estimator}. For ridge regression, we put the normal distribution $\mathcal{N} (0, \sigma^2\lambda^{-1} I)$ on the regression coefficients $\beta$. We estimate hyperparameters $\mu$, $\sigma^2$ and $\lambda$ by maximizing the marginal liklihood.  

We investigate the prediction accuracy of each procedure (EB, DPP, LDPP, GDPP, RIDGE, OLS). In particular, our interest is the robustness of the predictive performance of each method when the value of predictors $X$ in the training dataset and the test dataset are very different. In this setting, collinearity influences on the prediction.  In order to investigate the robustness, we divided the observations into two parts according to the values of predictors $X$. The first part is the candidate for the test dataset, and the second part is for the training dataset. To be precise, we divide the datasets as follows. Let $\tilde{X}$ be the design matrix before standardization and let $\tilde{x}_{ij}$ be the $(i,j)$ element of $\tilde{X}$. We calculate the mean $\bar{m}=(\tilde{m}_1,\ldots, \tilde{m}_p)^{\top}$ and the sample covariance matrix $\tilde{\Sigma}$ of $\tilde{X}$:
\begin{eqnarray}
\tilde{m}\ddef\bigg(\frac{1}{n}\sum_{i=1}^n\tilde{x}_{i1},\ldots, \frac{1}{n}\sum_{i=1}^n\tilde{x}_{ip}\bigg)^{\top}, \quad \tilde{\Sigma}\ddef \frac{1}{n-1}\tilde{A}^{\top}\tilde{A},
\end{eqnarray}
where
\begin{eqnarray}
\tilde{A} \ddef \tilde{X} - \bigg(\tilde{m}_11_n,\ldots, \tilde{m}_p1_n\bigg).
\end{eqnarray}
Using $\tilde{m}$ and $\tilde{\Sigma}$, we calculate the Mahalanobis distance from $\tilde{m}$ to each observation $\tilde{x}^k$ that is the $k$-th row of $\tilde{X}$. The furthest 10 observations from $\tilde{m}$ is assigned to the first part.  The second part consists of the remaining observations excluding the furthest $20$ (for the Air Pollution Data) or $50$ (for the Body Fat Data) observations from the  $\tilde{m}$. Note that 10 (for the Air Pollution Data) or $40$ (for the Body Fat Data) observations are not included in either part. This is because our purpose is to investigate the prediction accuracy when the values of predictors in the training and the test dataset are very different. For $l=1, 2, \ldots, n$ ($n=60$ for the Air Pollution Data and $n=100$ for the Body Fat Data), we randomly sample 1 observation $y_{\rm test}^{(l)}$ from the fist part and $m$ ($m=20$ for the Air Pollution Data and $m=30$ for the Body Fat Data) observations $X_{m}^{(l)}$ and $y_{m}^{(l)}$ from the second part. Then the prediction accuracy for each procedure is evaluated by the absolute loss function:
\begin{eqnarray}
|y_{\rm test}^{(l)}-\hat{y}_l|,
\end{eqnarray}
where $\hat{y}_l$ is the prediction value for $y_{\rm test}^{(l)}$ based on $X_{m}^{(l)}$ and $y_{m}^{(l)}$.

\subsection{Air Pollution Data}\label{airpollution}
We apply our methods to the Air Pollution Data which is a widely known collinear dataset. The Air Pollution Data was originally analyzed by \cite{McDonaldSchwing}. The data consists of daily mortality rates in $60$ Standard Metropolitan Statistical Areas of the USA,  and $15$ predictors. The dataset is available from R package SMPracticals \citep{airpollution}. 

In order to reduce the computational burden of estimating the hyperparameters in Bayesian variable selection (EB, DPP, LDPP and GDPP), we select 10 important predictors by least angle regression \citep{EfronEtAl} beforehand. To be precise, for $l=1,2,\ldots, 60$, we select 10 predictors $x_{j_1},\ldots, x_{j_{10}}$ by least angle regression and hyperparameters are estimated by maximizing the sum
\begin{eqnarray}\label{approximation_of_marginal}
\sum_{\gamma_{j_1}=\{0,1\}}\sum_{\gamma_{j_2}=\{0,1\}}\cdots \sum_{\gamma_{j_{10}}=\{0,1\}}p(y_n|\gamma, \xi)p(\gamma|\xi),
\end{eqnarray}
where $\xi$ denotes all hyperparameters to be estimated. Each evaluation of the type II likelihood $p(y_n|\xi)$ needs to sum of $p(y_n,\gamma|\xi)p(\gamma|\xi)$ $2^p$ times. Since this computation is a heavy task even if $p$ is moderately large, we approximate the marginal likelihood $p(y_n|\xi)$ by the partial sum \eqref{approximation_of_marginal}.

\begin{figure}[tb]
\begin{minipage}[c]{0.48\hsize}
\centering
\includegraphics[width=6cm, height=6cm]{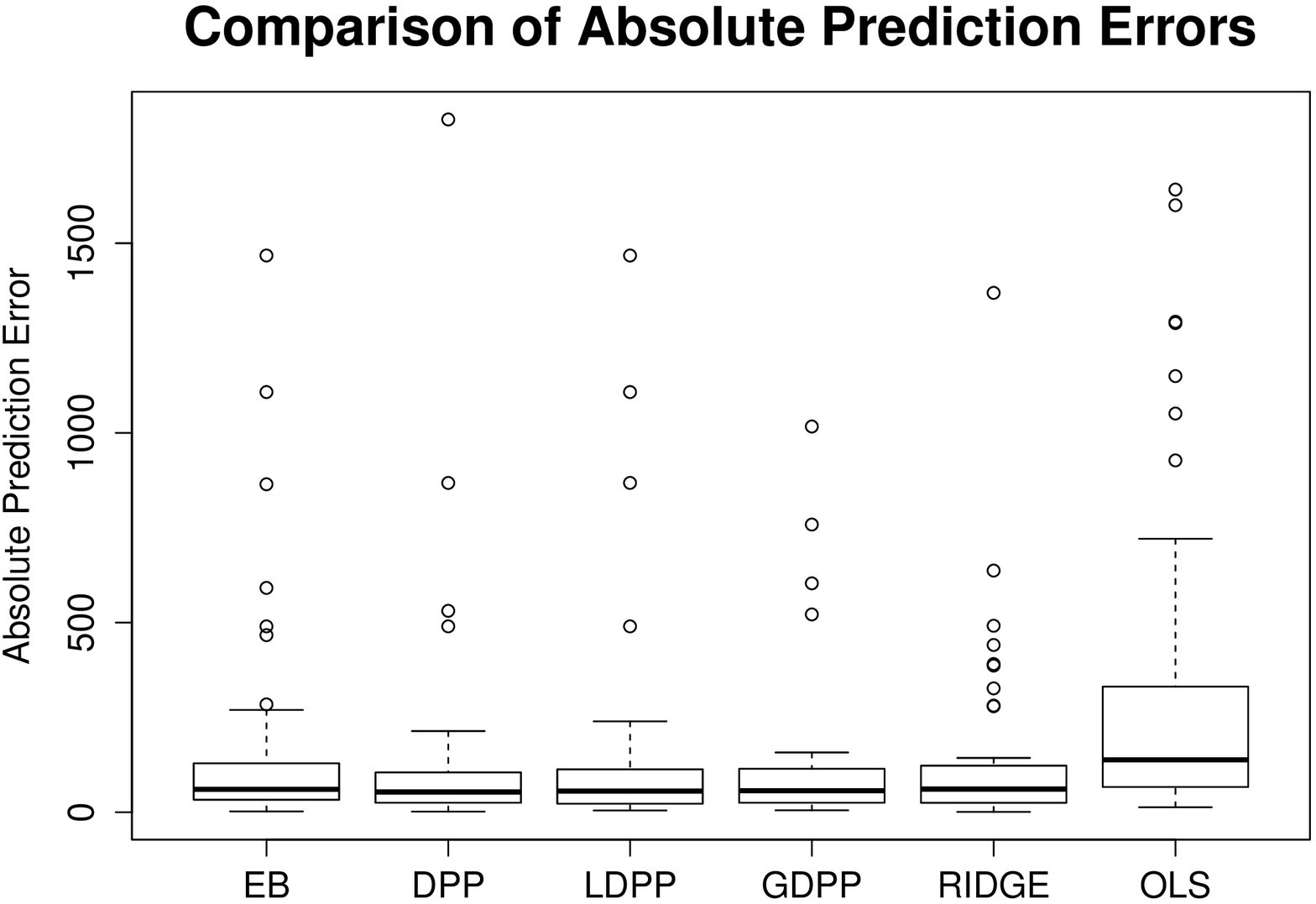}
\end{minipage}
\begin{minipage}[c]{0.48\hsize}
\centering
\includegraphics[width=6cm, height=6cm]{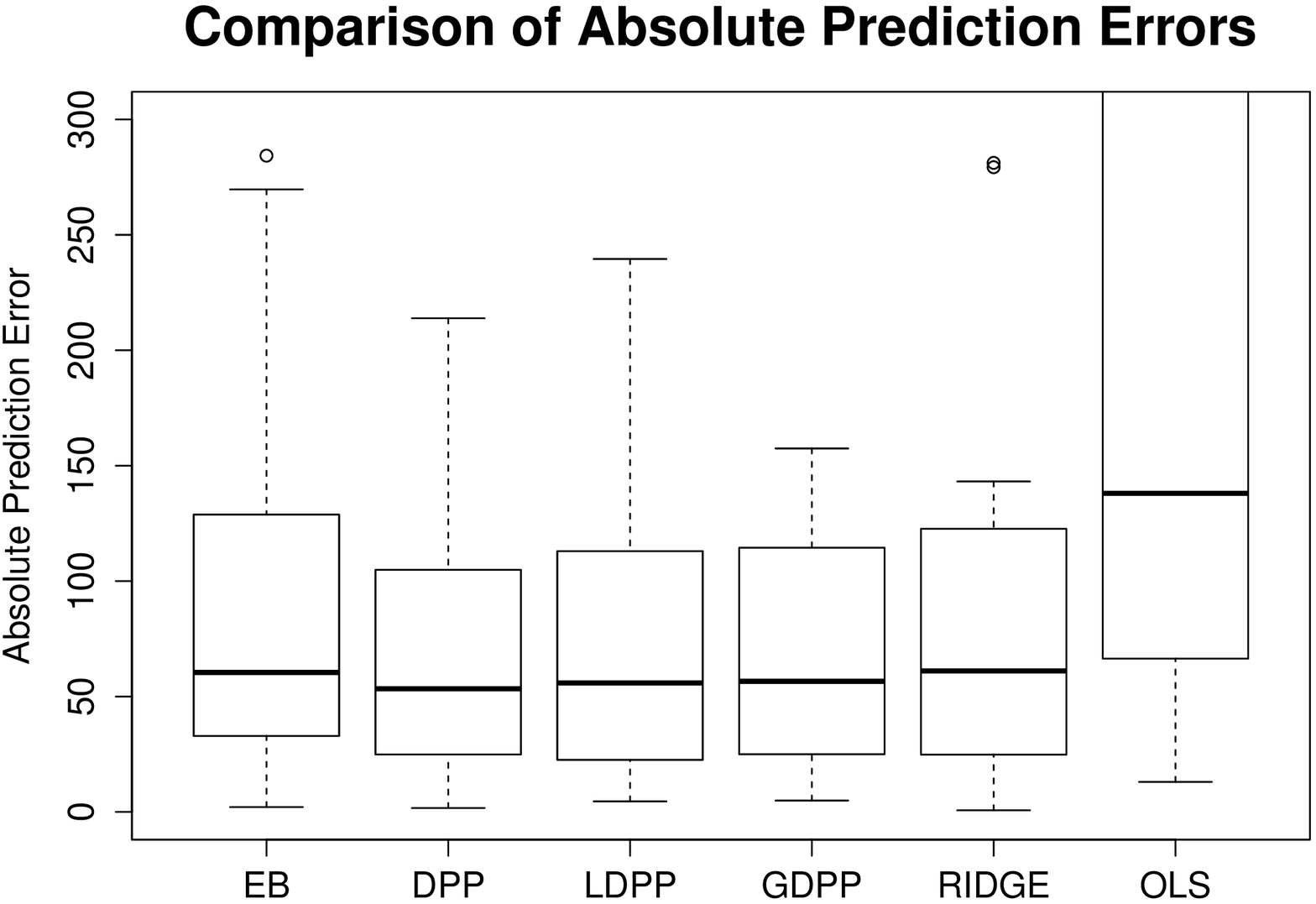}
\end{minipage}
\caption{Comparison of absolute prediction errors for EB, DPP, LDPP, RIDGE and OLS: The left panel shows the result of prediction by each procedure for the Air Pollution Data when we separate the dataset according to the Mahalanobis distances of $X$. The right panel shows an enlargement of the left panel.}
\label{result_of_prediction_airpollution_data}
\end{figure}

Figure \ref{result_of_prediction_airpollution_data} shows the result of prediction by each method. DPP, LDPP, GDPP and RIDGE perform better than EB and OLS. For comparison of the predictive performances of EB and DPP, we conduct a Wilcoxon signed rank test for paired samples. The alternative hypothesis is that DPP outperforms EB. As a result, the p-value of the test is $0.058$, and we consider DPP outperforms EB when the values of $X$ in the training and the test dataset are very different in the Air Pollution Data.

\subsection{Body Fat Data}\label{bodyfat}
Next, we report the Body Fat Data application. The dataset consists of estimates of the percentage of body fat determined by underwater weighing and 13 body circumference measurements for 252 men. To assess one's health, it is important to estimate the percentage of body fat. However, since accurate evaluation of body fat percentage is costly, we estimate the percentage from body circumference measurements such as neck circumference, ankle circumference and so on. Since we can estimate body fat percentages from body density by Siri's equation
\begin{eqnarray}
{\rm body~ fat} = 495/({\rm body~ density})-450,
\end{eqnarray}
estimating body density is enough to investigate body fat. Therefore we consider body density as the dependent variable. 
The dataset is available from Statlib (\url{http://lib.stat.cmu.edu/datasets/bodyfat}).

\begin{figure}
\centering
\includegraphics[width=8cm, height=8cm]{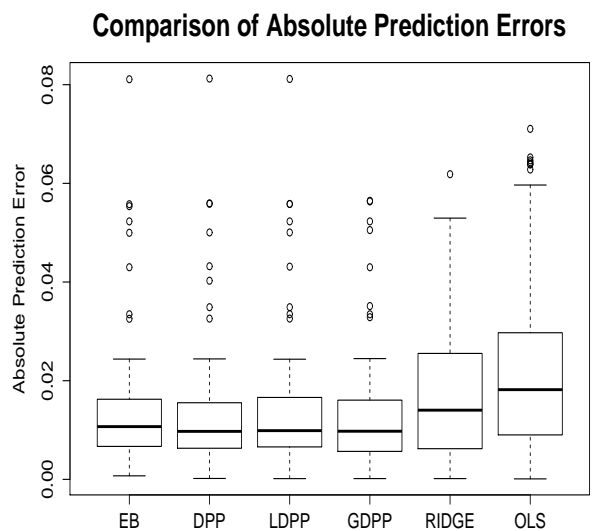}
\caption{Comparison of absolute prediction errors for EB, DPP, LDPP, RIDGE and OLS: The figure shows the result of prediction when we separate the dataset according to the Mahalanobis distances of $X$.}
\label{result_of_prediction_bodyfat_data}
\end{figure}

Figure \ref{result_of_prediction_bodyfat_data} shows the result of prediction by each procedure. Bayesian variable selection procedures (EB, DPP, LDPP, GDPP) yielded more accurate prediction than RIDGE and OLS. In order to compare the prediction accuracy of RIDGE and DPP, we conduct a Wilcoxon signed rank test for paired samples. The alternative hypothesis is that DPP outperforms RIDGE. As a result, the p-value of the test is $9.9\times 10^{-4}$, and we consider DPP outperforms RIDGE when the values of $X$ in the training and test dataset are very different in the Body Fat Data.

From Figure \ref{result_of_prediction_airpollution_data}, our methods (DPP, LDPP and GDPP) and RIDGE outperform EB and OLS for the Air Pollution Data. On the other hand, from Figure \ref{result_of_prediction_bodyfat_data}, our methods and EB outperform RIDGE and OLS. Therefore we conclude that the predictive performances of our methods are better than the other methods (EB, RIDGE and OLS) in the sense that the prediction by our methods are more accurate and robust. We consider the robustness comes from the repulsion property of DPPs. If the values of predictors $X$ in the training and the test dataset are very different, collinearity influences on prediction. In this setting, usual methods (EB, RIDGE and OLS) are inappropriate because such methods do not take into account for the correlations among predictors $X$. However, since DPP priors assign small prior probabilities to submodels including collinear predictors, prediction by our methods are robust and accurate. 

\section{Conclusion}
We considered Bayesian variable selection in linear regression, and proposed discrete determinantal point processes (DPPs) for prior distributions on model parameter $\gamma$. Since the proposed prior (DPP prior) assigns small probabilities to submodels including collinear predictors, collinear predictors are less likely to be selected simultaneously. In Section \ref{dpp}, we see that DPP prior is a generalization of the Bernoulli distribution that is used for $p(\gamma)$ in the method proposed by \cite{GeorgeFoster} (EB). Therefore our method is a generalization of  EB. From this point of view, we also proposed linear mixture DPP prior (LDPP) and geometric mixture DPP prior (GDPP) that bridge the Bernoulli distribution and DPP prior. 

In the numerical experiment, the estimators of regression coefficients constructed by our methods reduce the maximum risk more than the other estimators (EB, the ridge estimator (RIDGE) and the ordinary least squares estimator (OLS)) when collinearity is severe. We also apply our methods to the Air Pollution Data and the Body Fat Data. Our interest is the robustness of the predictive performance of each method when the value of predictors $X$ in the training dataset and the test dataset are very different. For the Air Pollution Data, proposed methods and RIDGE yielded more accurate prediction than EB and OLS. In addition, for the Body Fat Data, proposed methods and EB yielded more accurate prediction than RIDGE and OLS. From these results of the applications, we conclude that prediction of our methods are more accurate and robust comparing to the other methods (EB, RIDGE and OLS). We consider the robustness comes from the repulsion property of DPPs.

Finally, we give a future plan of this work. In the large $p$ small $n$ setting, we intend to use the proposed DPP priors, combining the stochastic search method proposed by \cite{KwonEtAl}. In this setting, since too many predictors exist and collinearity is severe, our methods will be efficient. 

%%%%references%%%%
\bibliography{ref_metr.bib}
\end{document}